\def\asca{{\it ASCA\/}}

\def\einstein{{\it Einstein\/}}

\def\rosat{{\it ROSAT\/}}
\def\sax{{\it BeppoSAX\/}}
\def\xmm{{\it XMM\/}}

\def\cso{CSO~755}

\def\aox{{$\alpha_{\rm ox}$}}

\def\ltsima{$\; \buildrel < \over \sim \;$}
\def\simlt{\lower.5ex\hbox{\ltsima}}
\def\gtsima{$\; \buildrel > \over \sim \;$}
\def\simgt{\lower.5ex\hbox{\gtsima}}
        
\documentstyle[12pt,psfig,aaspp4]{article}

\begin{document}

% ------------------------------------------------------------------

\title{X-rays from the Highly Polarized Broad Absorption Line QSO \cso}

% ------------------------------------------------------------------

\author{W.~N.~Brandt,\footnotemark\
A.~Comastri,\footnotemark\
S.~C.~Gallagher,$^1$ 
R.~M.~Sambruna,$^1$
Th.~Boller\footnotemark\ and 
A.~Laor\footnotemark}

% \affil{Department of Astronomy and Astrophysics,
% 525 Davey Laboratory,
% Pennsylvania State University,
% University Park, PA 16802}

% \author{Boller, Comastri, Laor}
% \affil{Other addresses}

% \author{A. Laor}
% \affil{Physics Department,
% Technion,
% Haifa 32000 Israel}

% ------------------------------------------------------------------

\footnotetext[1]{Department of Astronomy and Astrophysics,
525 Davey Laboratory, Pennsylvania State University, University Park, PA 16802}

\footnotetext[2]{Osservatorio Astronomico di Bologna, via Ranzani 1, I-40127 Bologna, Italy}

\footnotetext[3]{Max-Planck-Institut f\"ur Extraterrestrische Physik, 85748 Garching, Germany}

\footnotetext[4]{Physics Department, Technion, Haifa 32000, Israel}

\begin{abstract}
We present results from a \sax\ observation of the BAL~QSO \cso, 
observed as part of our program to investigate the X-ray properties of 
highly polarized BAL~QSOs. \cso\ is clearly detected by the \sax\ MECS, 
making it the highest redshift ($z=2.88$) and most optically luminous
($M_{\rm V}=-27.4$) BAL~QSO seen in 
X-rays. It is detected in several energy bands 
including the rest-frame 21--39~keV band, but we are only able to place 
loose constraints upon its X-ray spectral shape. Our X-ray detection 
is consistent with the hypothesis that the BAL~QSOs with high optical 
continuum polarization tend to be the X-ray brighter members of the 
class. We examine a scattering interpretation of a polarization/X-ray 
flux connection, and we discuss the data needed to prove or refute 
such a connection. We also discuss a probable \rosat\ detection of \cso. 
The observed-frame 2--10~keV flux from \sax\ 
($1.3\times 10^{-13}$~erg~cm$^{-2}$~s$^{-1}$) is high enough to allow 
\xmm\ spectroscopy, and studies of iron~K line emission should prove 
of particular interest if a large amount of scattered X-ray flux
is present. 
% \vspace*{1.8 in}
\end{abstract}

% Somewhat surprisingly, its \aox\ value appears to be consistent with 
% those of typical radio-quiet QSOs. 

% ------------------------------------------------------------------


% ------------------------------------------------------------------

\section{Introduction}

The ejection of matter at moderate to high velocities is a common and 
perhaps universal phenomenon of Quasi-Stellar Objects (QSOs). One of 
the main manifestations of QSO outflows is the blueshifted UV Broad 
Absorption Lines (BALs) seen in $\sim 10$\% of optically selected QSOs, 
the BAL~QSOs (e.g., Weymann 1997). X-ray spectroscopy of BAL~QSOs is 
potentially important for studying their outflows and nuclear geometries, 
but the study of BAL~QSOs in the X-ray regime has not yet matured, 
largely due to low X-ray fluxes (e.g., Green \& Mathur 1996; 
Gallagher et~al. 1999). Only $\approx 9$ BAL~QSOs have been 
detected in X-rays at present. The current data suggest
that the X-ray emission from BAL~QSOs suffers from significant
intrinsic absorption, with many BAL~QSOs having absorption column 
densities $\simgt$~(1--5)$\times 10^{23}$~cm$^{-2}$. Optical 
brightness is {\it not\/} a good predictor of X-ray brightness for
BALQSOs; some optically faint BAL~QSOs have been clearly 
detected (e.g., PHL~5200; $V=18.1$) while some of the optically brightest 
(e.g., PG~$1700+518$; $V=15.1$) remain undetected in deep 0.1--10~keV 
observations. In the limited data available at present, however, 
there is a suggestion that the BAL~QSOs with 
high ($\simgt 2$\%) optical continuum polarization 
{\it may\/} be the X-ray brighter members of the class (see \S4 of 
Gallagher et~al. 1999). A polarization/X-ray flux connection, 
if indeed present, would provide a clue about the geometry 
of matter in BAL~QSO nuclei (see \S3). 

To improve understanding of the X-ray properties of BAL~QSOs
and examine the possible polarization/X-ray flux connection, 
we have started a program to observe highly polarized 
BAL~QSOs in X-rays. An excellent target for this program was the 
Case Stellar Object 755 (\cso; $z=2.88$; Sanduleak \& Pesch 1989), 
which has $V=17.1$ (e.g., Barlow 1993) and is a representative, 
`bona-fide' BAL~QSO in terms of its luminosity and UV absorption 
properties (e.g., Glenn et~al. 1994). Its continuum polarization 
is high ($\approx$~3.8--4.7\%; only 8/53 BAL~QSOs studied by
Schmidt \& Hines 1999 had $>2$\%) and rises to the blue. 

% Here we report the results and interpretation of our 
% recent \sax\ (Boella et~al. 1997) observation of \cso. 

We adopt $H_0=70$~km~s$^{-1}$ Mpc$^{-1}$ and $q_0=\frac{1}{2}$. 
The Galactic neutral hydrogen column density towards \cso\ is 
$(1.6\pm 0.4)\times 10^{20}$ cm$^{-2}$ (Stark et~al. 1992). 

% ------------------------------------------------------------------

\section{Observations, Analysis and Results}

We observed \cso\ with \sax\ (Boella et~al. 1997) on 
1999~Feb~2. We will focus on the results 
from the Medium-Energy Concentrator Spectrometers
(MECS; 1.8--10~keV; 35.2~ks exposure) and Low-Energy Concentrator 
Spectrometer (LECS; 0.1--4~keV; 12.7~ks exposure), since the data 
from the other instruments are not useful for such a faint source.
Our energy coverage corresponds to 0.4--39~keV in the 
rest frame. The observation 
went smoothly, and the resulting data were
processed with Version~1.2 of the \sax\ Science Data Center (BSDC)
pipeline. We have adopted the standard reduction methods 
recommended by the BSDC (Fiore, Guainazzi \& Grandi 1999), and we 
do not observe any irregular background variability. 

The screened events resulting from the above reduction were 
analyzed using {\sc xselect}. We 
made full-band images for each of the detectors as well as
combined MECS2+MECS3 images. An X-ray source consistent
with the precise optical position of \cso\ is detected with
high statistical significance in our MECS2, MECS3 and MECS2+MECS3 
images (e.g., Figure~1), but it is not detected by the LECS.
Given the observed flux (see below), the probability of a
confusing source is $\simlt 5\times 10^{-3}$, and no
particularly suspicious sources are found in the Palomar Optical 
Sky Survey or the {\sc ned/simbad} catalogs. 
To determine MECS count rates, we have used a $3^\prime$-radius 
circular source cell centered on the X-ray centroid. For
background subtraction, we use five $3^\prime$-radius
circular cells near \cso\ (we have not used an
annulus because a weak nearby source would fall inside the
annulus). We have corrected for energy-dependent vignetting of the 
background following \S3.1.5 of Fiore et~al. (1999). 
In the MECS2+MECS3 full-band (1.8--10~keV) image, we detect 
$54.3\pm 14.3$ counts from \cso\ for a MECS2+MECS3 count rate of
$(1.5\pm 0.4)\times 10^{-3}$~count~s$^{-1}$. The LECS
$3\sigma$ upper limit on the 0.1--1.8~keV count rate is
$<1.7\times 10^{-3}$~count~s$^{-1}$ (computed using a circular
source cell with a $5^\prime$ radius). 

While we do not have enough photons for spectral fitting, 
we have analyzed MECS2+MECS3 images in three observed-frame
energy bands to place crude constraints 
on spectral shape: 
1.8--3~keV (band~1; channels 40--66), 
3--5.5~keV (band~2; channels 67--120), and 
5.5--10~keV (band~3; channels 121--218).  
\cso\ is detected in all bands, 
although with varying degrees of statistical significance. 
We give the corresponding count rates in Table~1, and 
the Poisson probabilities of false detections in 
bands~1, 2 and 3 are 
$6.8\times 10^{-5}$,
$4.8\times 10^{-3}$ and
$2.8\times 10^{-2}$, respectively.  
The detection in band~3 (21--39~keV in the rest frame) is notable. 

To compare the observed spectral shape with spectral models, we have 
employed a band-fraction diagram similar to those used in studies of 
the diffuse soft X-ray background (e.g., see \S5 of Burstein et~al. 1977). 
We first consider a simple power-law model with photon index, 
$\Gamma=$~1.7--1.9 (a typical, representative range for radio-quiet QSOs;
e.g. Reeves et~al. 1997), and 
neutral absorption at $z=2.88$. For this model, Figure~2 shows 
that column densities less than 
$\approx 7\times 10^{23}$~cm$^{-2}$ are most consistent with our data. 
Alternatively, for small column densities, values of $\Gamma$
down to $\approx 0.8$ are most consistent with our data 
(i.e. the spectrum could be as flat as that for a 
`reflection-dominated' source). 
Incorporating the LECS upper limit into similar analyses does not 
significantly tighten our constraints. 

If we consider a $\Gamma=1.9$ power-law model with 
the Galactic column density, we calculate an observed-frame 2--10~keV 
flux of $1.3\times 10^{-13}$~erg~cm$^{-2}$~s$^{-1}$, corresponding
to a rest-frame 7.8--39~keV luminosity of $4.0\times 10^{45}$~erg~s$^{-1}$. 
These two quantities are relatively insensitive to the internal column 
density for $N_{\rm H}<5\times 10^{23}$~cm$^{-2}$. If we extrapolate
this model into the rest-frame 2--10~keV band, the luminosity is 
$3.4\times 10^{45}$~erg~s$^{-1}$. 

We have also calculated \aox\ (the slope of a hypothetical
power law between 3000~\AA\ and 2~keV in the rest frame), since 
this parameter can be used as a statistical predictor of the 
presence of X-ray absorption (e.g., Brandt, Laor \& Wills 1999). 
We calculate the rest-frame 3000~\AA\ flux density using the  
observed-frame 7500~\AA\ flux density of Glenn et~al. (1994)
and a continuum spectral index of $\alpha=0.5$. 
The rest-frame flux density at 2~keV 
is more difficult to calculate since we do not 
have strong constraints on X-ray spectral shape or a \sax\ detection
at ${2~{\rm keV}\over (1+z)}=0.52$~keV (although see our discussion
of the \rosat\ data below). If we 
normalize a $\Gamma=1.9$ power-law model with Galactic 
absorption to the rest frame 7--39~keV count rate (corresponding to
1.8--10~keV in the observed frame), we calculate $\alpha_{\rm ox}=1.58$. 
Of course, this $\alpha_{\rm ox}$ value is really telling us about the 
rest-frame 7--39~keV emission rather than a directly measured flux 
density at 2~keV. 

% We calculate the rest-frame 3000~\AA\ flux density from 
% the $R$ magnitude of 16.9 (McMahon \& Irwin 1992). 

We have searched for any \einstein, \rosat\ or \asca\ pointings 
that serendipitously contain \cso, but unfortunately there is 
none.
%
%  (\cso\ lies just outside the \rosat\ field rf200808n00). 
%
We have also analyzed the data from the \rosat\ All-Sky Survey (RASS). 
\cso\ was observed for 939~s during the RASS between 1990~Dec~31 and
1991~Jan~4 (a relatively long RASS exposure; see Figure~2 of
Voges et~al. 1999). There appears to be an $\approx 7$-photon 
enhancement over the average background at the position of \cso. 
Comparative studies of RASS and pointed data show that $\approx$~90\% 
of such 7-photon RASS sources are real X-ray sources rather than 
statistical fluctuations, and \cso\ is included in the 
Max-Planck-Institut f\"ur Extraterrestrische Physik RASS
faint source catalog (Voges et~al., in preparation) with a 
likelihood of 11 (see Cruddace, Hasinger \& Schmitt 1988). 
However, to be appropriately cautious we shall treat the 
probable RASS detection as tentative.
The probable RASS detection corresponds to a 
vignetting-corrected flux in the observed 0.1--2.4~keV band 
of $\approx~1.1\times 10^{-13}$~erg~cm$^{-2}$~s$^{-1}$
(for a power-law model with $\Gamma=1.9$ and the Galactic
absorption column). Given the relative effective areas and imaging 
capabilities of the \rosat\ PSPC and \sax\ LECS, a RASS detection 
is consistent with the LECS upper limit given in Table~1 
(see Figure~2 of Parmar et~al. 1999). Provided there is not 
substantial intrinsic X-ray absorption below the MECS band, 
the relative RASS and MECS fluxes are entirely plausible. 
If we use the \rosat\ flux to normalize a $\Gamma=1.9$ power law
with Galactic absorption, we calculate $\alpha_{\rm ox}=1.62$.
If \rosat\ has indeed detected \cso, the \rosat\ band has the
advantage that it directly constrains the rest-frame 2~keV 
flux density. 

% ------------------------------------------------------------------ 

\section{Discussion and Conclusions}

Our \sax\ and probable \rosat\ detections 
of \cso\ make it the highest redshift as well as the
most optically luminous BAL~QSO detected in X-rays. It was 
selected for study not based upon high optical 
flux but rather based on its high (observed-frame)
optical continuum polarization
(3.8--4.7\%; hereafter OCP), and it is X-ray brighter than several 
other BAL~QSOs that have $\approx$~4--6 times its $V$-band flux 
(compare with Gallagher et~al. 1999). While its higher X-ray flux 
could partially result from the higher redshift providing
access to more penetrating X-rays (i.e. a `negative $K$-correction'), 
there is also suggestive evidence that the BAL~QSOs with high OCP 
may be the X-ray brighter members of the class. 

We have investigated the OCP percentages of the 10 BAL~QSOs 
(including \cso) with reliable X-ray detections using the data 
from Berriman et~al. (1990), Hutsem\'ekers, Lamy \& Remy (1998), 
Ogle (1998) and Schmidt \& Hines (1999). The OCP percentages have 
a mean of $2.28\pm 0.28$, a standard deviation of 0.88, and a median 
of 2.24. These values indeed place the X-ray detected BAL~QSOs 
toward the high end of the BAL~QSO OCP distribution function
(compare with \S2 of Schmidt \& Hines 1999). At present, however,
our nonparametric testing is unable to prove 
that the X-ray detected BAL~QSOs have higher OCPs than those that
are undetected in sensitive X-ray observations. This is due
to small sample sizes as well as concerns about possible 
secondary correlations and observational biases. Many of the 
BAL~QSOs with high-quality X-ray data have been
observed because they have exceptional properties 
(e.g., low-ionization absorption, extreme Fe~{\sc ii} 
emission), and thus the currently available sample is not 
necessarily representative of the population as a whole.
In addition, the current X-ray and polarization observations of 
BAL~QSOs span a wide range of rest-frame energy/wavelength bands 
due to redshift and instrumentation differences (redshifts for the 
10 X-ray detected BAL~QSOs run from $z=$~0.042--2.88). 
At higher redshifts one samples harder X-rays that are less
susceptible to absorption. Also at higher redshifts, observed-frame 
OCP measurements tend to sample shorter wavelengths, and many QSOs
show polarization that rises towards the blue. 
Systematic X-ray and polarimetric observations of uniform, 
well-defined BAL~QSO samples are needed to examine this issue better. 

A polarization/X-ray flux connection could be physically 
understood if the direct lines of sight into the X-ray nuclei of 
BAL~QSOs were usually blocked by extremely thick matter 
($\gg 10^{24}$~cm$^{-2}$). In this case, we could only see
X-rays when there is a substantial amount of electron
scattering in the nuclear environment by a `mirror' of moderate
Thomson depth.\footnote{Ogle (1998) suggests that there is a 
large range of mirror optical depths among the BAL~QSO
population.} The scattering would provide a periscopic, indirect 
view into the compact X-ray emitting region while also polarizing 
some of the more extended optical continuum emission (see 
Figure~3). Measured X-ray column densities would then provide 
information only about the gas along the {\it indirect\/} line 
of sight. 
For \cso, the X-ray scattering medium would need to be located 
on fairly small scales ($\simlt$ a few light weeks) to satisfy 
the spectropolarimetric constraints of Glenn et~al. (1994) and 
Ogle (1998). These show that the material scattering the 
optical light is located at smaller radii than both the 
Broad Line Region (BLR) and BAL region.

Our calculations in \S2 give an \aox\ value of $\approx$~1.6,
although our only direct constraint on the rest-frame 2~keV flux
density is via the probable \rosat\ detection. Our \aox\ value
is entirely consistent with those of typical radio-quiet
QSOs (compare with Figure~1 of Brandt, Laor \& Wills 1999), and it
is smaller than those of many BAL~QSOs (e.g. Green \& Mathur 1996). 
A `normal' \aox\ value would appear somewhat surprising in the
context of the scattering model of the previous paragraph, since
one would expect the X-ray flux level to be reduced if the
direct line of sight is blocked. However, there is enough 
dispersion in the \aox\ distribution that the observed value
of \aox\ does not cause a serious problem, provided the 
scattering is efficient. The scattering mirror would need to
subtend a fairly large solid angle (as seen by the compact 
X-ray source) and have a moderate Thomson depth 
(say $\tau_{\rm T}\sim 0.3$). In addition, there may be
`attenuation' at 3000~\AA\ (in the sense of \S2 of Goodrich 1997) 
that helps to flatten \aox. 

Finally, we note that \cso\ has a high enough X-ray flux to allow 
moderate quality X-ray spectroscopy and variability studies
with \xmm. It is currently scheduled for a 5~ks \xmm\ 
observation, but this is an inadequate exposure time for 
such work. A longer \xmm\ exposure 
would allow a study of any iron~K spectral features, and  
the high redshift of \cso\ moves the iron~K complex 
right to the peak of the \xmm\ EPIC spectral response. If we are 
viewing a large amount of scattered X-ray flux in \cso\ and other
high polarization BAL~QSOs, then narrow iron~K lines with large 
equivalent widths may be produced via fluorescence and resonant 
scattering (as for the much less luminous Seyfert~2 galaxies; e.g., 
Krolik \& Kallman 1987). Such lines could allow direct detection 
of the X-ray scattering medium, and line energies and 
blueshifts/redshifts would constrain the ionization state and 
dynamics of the mirror. We would also not expect rapid ($\simlt 1$~day) 
and large-amplitude X-ray variability if most of the X-ray flux 
is scattered. 

% ALSO IRON EDGE AND COMPTON HUMP

% ------------------------------------------------------------------

\acknowledgments

We thank J. Halpern, J. Nousek, W. Voges and B. Wills for helpful discussions, 
and we thank H. Ebeling for the use of his {\sc idl} software. 
We acknowledge the support of 
NASA LTSA grant NAG5-8107 (WNB),
Italian Space Agency contract ASI-ARS-98-119 and MURST grant Cofin-98-02-32 (AC),  
NASA grant NAG5-4826 and the Pennsylvania Space Grant Consortium (SCG), and 
the fund for the promotion of research at the Technion (AL).

% ------------------------------------------------------------------

\newpage

% ---------------

\begin{figure}
\epsscale{0.5}
\plotfiddle{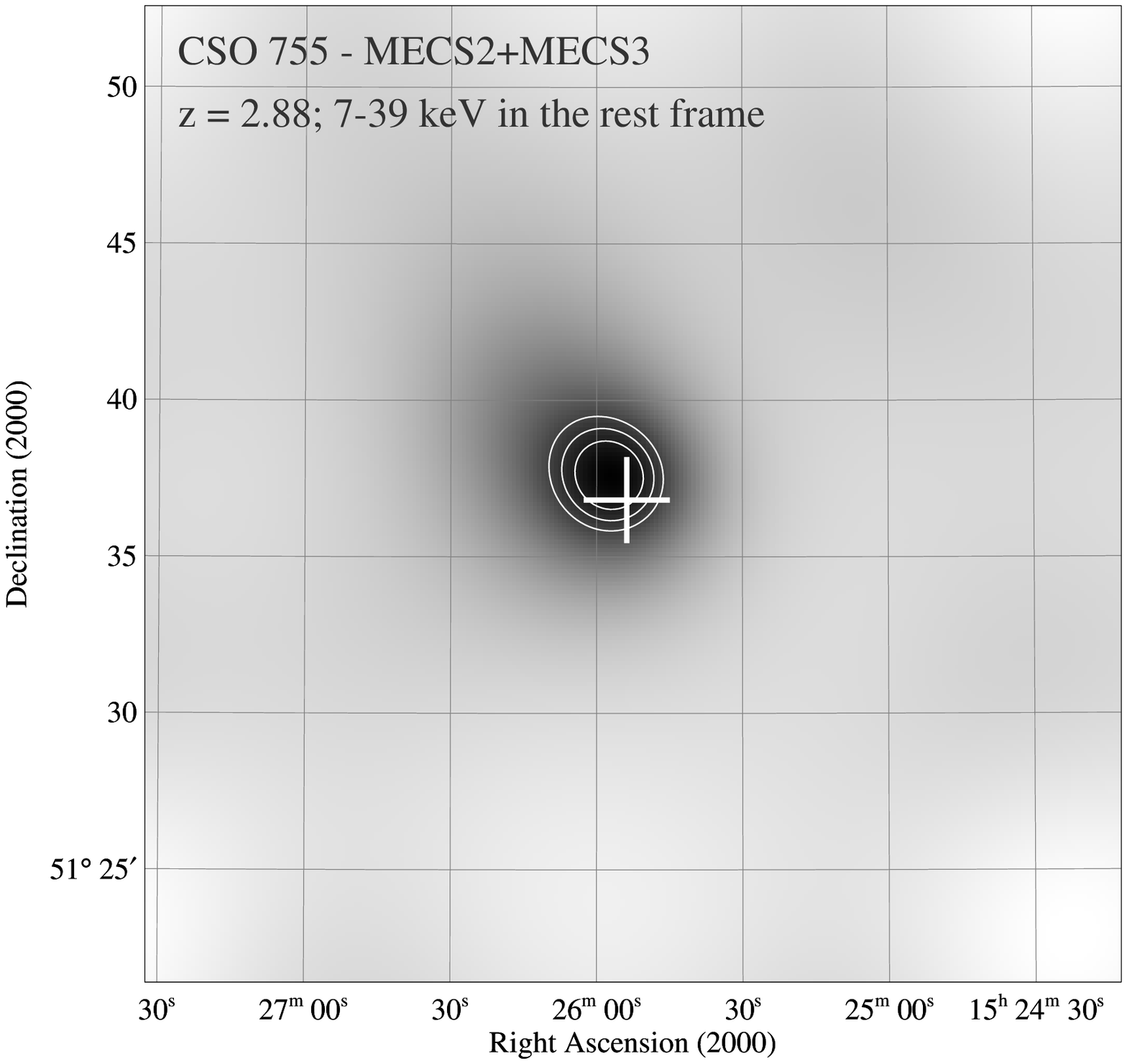}{360pt}{0}{70}{70}{-240}{-100}
% \vspace{-1.5in}
\caption{MECS2+MECS3 image of \cso\ made using the 
full-band (1.8--10~keV) data. This energy range 
corresponds to 7--39~keV in the rest frame. The image 
has been adaptively smoothed using the algorithm
of Ebeling, White \& Rangarajan (1999). The contours are 
drawn at 85.1, 89.4 and 93.7\% of the maximum pixel value, 
and the white cross marks the precise optical position of
\cso.
\label{fig1}}
\end{figure}

\clearpage

% ---------------

\begin{figure}
\epsscale{0.5}
\plotfiddle{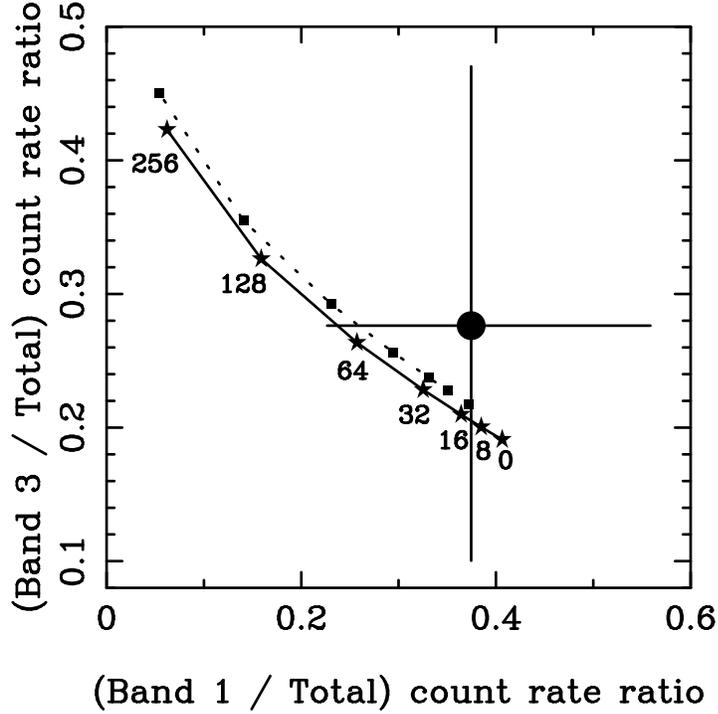}{360pt}{0}{70}{70}{-240}{-100}
\caption{Band-fraction diagram for \cso. 
The large solid dot with error bars shows the position of \cso. 
The error bars have been computed using the `numerical method' 
described in \S1.7.3 of Lyons (1991) and roughly correspond to
$1\sigma$. 
The stars and solid curve show the locus of a power-law model 
with photon index $\Gamma=1.9$ and varying 
amounts of $z=2.88$ neutral absorption ($N_{\rm H}$). 
The squares and dotted curve show the same for a power-law 
model with $\Gamma=1.7$. Absorption 
increases towards the left, and the individual symbols 
along each curve correspond to $N_{\rm H}$ values of
(0, 8, 16, 32, 64, 128 and 256)$\times 10^{22}$~cm$^{-2}$. The stars
for the $\Gamma=1.9$ case have been labeled in units of $10^{22}$~cm$^{-2}$. 
\label{fig2}}
\end{figure}

\clearpage

% ---------------

\begin{figure}
\epsscale{0.5}
\plotfiddle{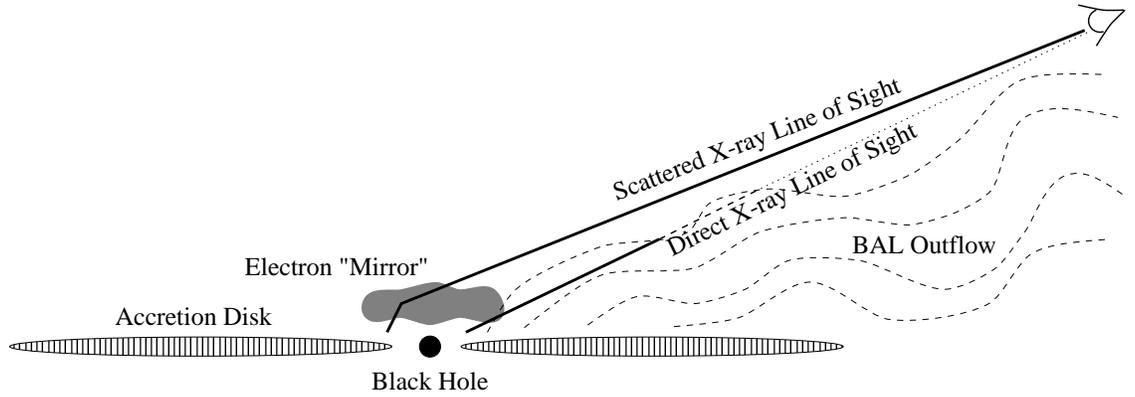}{360pt}{-90}{60}{60}{-200}{+500}
\caption{
A schematic model showing the possible X-ray lines of sight in a highly
polarized BAL~QSO. X-rays which would travel directly to the observer
are strongly absorbed by extremely thick matter ($\gg 10^{24}$~cm$^{-2}$). 
X-rays that are scattered by a compact electron `mirror' of moderate
Thomson depth are able to reach the observer with significantly less
absorption and thus dominate the observed flux. While we have drawn the
BAL outflow to be the X-ray absorbing matter, we note that this has not 
been demonstrated to be the case. 
\label{fig3}}
\end{figure}

\clearpage

% ---------------

\newpage

% ------------------------------------------------------------------

\begin{table*}
\small
\renewcommand{\arraystretch}{1.3}
\renewcommand{\tabcolsep}{2mm}
\begin{center}
TABLE 1

{\sc Count Rates in the Observed Frame}
\vspace{1mm}

\begin{tabular}{ll}
\hline \hline
Instrument \&    & Count Rate /                \\
Energy Band      & (10$^{-4}$ Count s$^{-1}$)  \\
\hline
\sax\ LECS                             &              \\
\hspace{0.07 in} 0.1--1.8~keV          & $<17$        \\
\sax\ MECS2+MECS3                         &              \\
\hspace{0.07 in} 1.8--3~keV (band~1)      & $5.8\pm 1.9$ \\
\hspace{0.07 in} 3--5.5~keV   (band~2)    & $5.4\pm 2.5$ \\
\hspace{0.07 in} 5.5--10~keV  (band~3)    & $4.3\pm 2.6$ \\
\hspace{0.07 in} 1.8--10~keV  (full band) & $15\pm 4$    \\
\hline
\end{tabular}
\vspace{1mm}

\begin{minipage}{17.5cm}
% $^{\rm a}$All values are for $h=0.5$. 
\end{minipage}
\end{center}
\vspace*{-2mm}
\end{table*}

\clearpage

% ------------------------------------------------------------------

\begin{figure}
\epsscale{0.5}
\plotfiddle{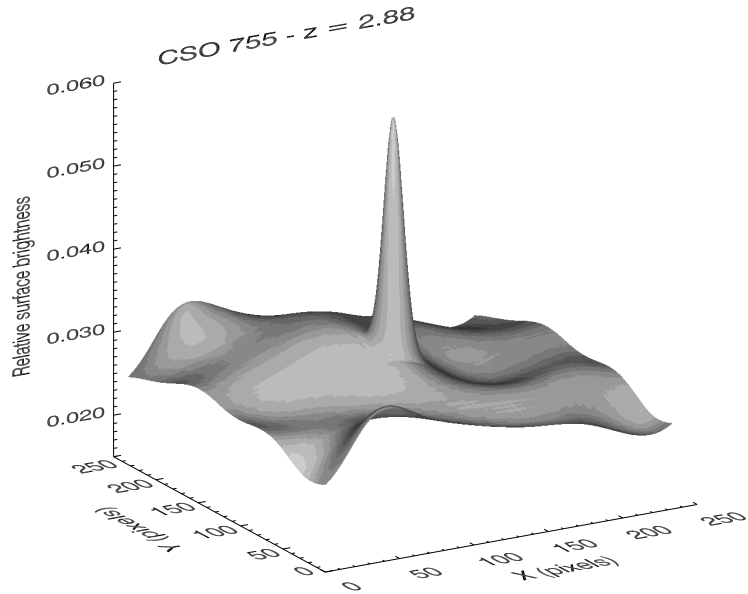}{360pt}{0}{70}{70}{-240}{-100}
% \vspace{-1.5in}
\caption{Three-dimensional representation of the 5.5--10~keV (band~3) 
image from MECS2+MECS3. The image has been adaptively smoothed, and
it corresponds to 21--39~keV in the rest frame. The peak in 
the center of the image is \cso. This is an unofficial `bonus' 
figure that will not appear in the ApJL publication.     
\label{fig4}}
\end{figure}

% \clearpage

% ---------------

\end{document}